\documentclass[twocolumn,aps,pr,superscriptaddress,preprintnumbers,nofootinbib,10pt]{revtex4-2}
\usepackage{amsmath,amssymb}
\usepackage[dvipdf,dvips]{graphicx}
\usepackage{color}
\usepackage{hyperref}
\usepackage{url}
\usepackage{slashed}
\usepackage{subfigure}
\usepackage[usenames,dvipsnames]{xcolor}
\usepackage{amsmath}
\usepackage{amsfonts}
\usepackage{float} 
\usepackage{amssymb}
\usepackage{epsfig}
\usepackage{graphics}
\usepackage{euscript}
\usepackage{slashed}
\usepackage{epstopdf}
\usepackage[utf8]{inputenc}
\allowdisplaybreaks
\usepackage[normalem]{ulem}
\usepackage{pifont}
\usepackage{dsfont}
\usepackage{MnSymbol}
\usepackage{verbatim}
\usepackage{graphicx}
\usepackage{latexsym}
\usepackage{courier}

\usepackage{tikz-feynman}

\hypersetup{
colorlinks=true,
citecolor=blue,
citebordercolor=red,
linktoc=all,
linkcolor=blue,
urlcolor=blue
}

\def \s{\mathsf{s}}

\def \and{\textmd{and}}

\def \be{\begin{equation}}
\def \ee{\end{equation}}

\def \bea{\begin{eqnarray}}
\def \eea{\end{eqnarray}}

\newbox{\ORCIDicon}
\sbox{\ORCIDicon}{\large
                  \includegraphics[width=0.8em]{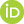}}

\begin{document}

\title{On the consistency of Lorentz-Violating Yang-Mills theories:\\
Gauge invariance and non-perturbative effects}

\author{Antonio D. Pereira\,\href{https://orcid.org/0000-0002-6952-2961}{\usebox{\ORCIDicon}}} \email{adpjunior@id.uff.br}
\affiliation{Instituto de F\'isica, Universidade Federal Fluminense, Campus da Praia Vermelha, Av. Litor\^anea s/n, 24210-346, Niter\'oi, RJ, Brazil}

\begin{abstract}
Previous investigations on the renormalizability properties of Lorentz-violating Yang-Mills (LVYM) theories in the Landau gauge have pointed out the necessity of the inclusion of a mass-like term for the gauge fields.  If one aims at generalizing the theory to a more complicated gauge, such a mass-like term can bring severe issues regarding gauge-dependence of correlation functions of observables. We propose a Lorentz-violating Yang-Mills theory supplemented by a gauge-invariant mass term which generalizes the model in the Landau gauge. We discuss the foundational aspects of the model and highlight how it simplifies in the Landau gauge. Following the recent literature in pure Yang-Mills theories, we remark how the proposed massive extension of LVYM theories can be used as an effective model to access infrared properties within perturbation theory. Finally, we present a BRST-invariant formulation of the so-called Refined Gribov-Zwanziger action in the presence of Lorentz-violating terms in linear covariant gauges and the underlying tree-level gauge-field propagator which is enriched by the non-perturbative information carried by the elimination of Gribov copies. 
\end{abstract}

\maketitle

%-------------------------------------------------------
\section{Introduction \label{Sec:Intro}}
%-------------------------------------------------------

Yang-Mills (YM) theories are the foundational pillar of the Standard Model of Particle Physics (SM). Besides their remarkable capacity to capture the phenomenology of Particle Physics, YM theories enjoy several important structural properties such as asymptotic freedom \cite{Politzer:1973fx,Gross:1973id}. This means that for sufficiently high energies, YM theories become weakly coupled ensuring that perturbation theory can be safely employed. This has enabled a myriad of explicit computations with very good accuracy. Nevertheless, towards the infrared, YM theories are strongly coupled. Perturbation theory breaks down and sophisticated non-perturbative techniques are necessary in order to properly describe the dynamics of the theory in this regime. A proper understanding of YM theories in their strongly coupled region is a challenging open problem that has been investigated for many years through several different approaches. In particular, this should provide a consistent comprehension of color confinement from first principles, see, e.g., \cite{Greensite:2011zz,Brambilla:2014jmp}. 

Much of the progress in the construction of modern quantum field theory is based on symmetry principles. In particular, given certain symmetry restrictions together with power-counting renormalizability and locality, one is able to almost fix completely an action to be used in a given model. In particular, gauge and Lorentz symmetries play a pivotal role in the formulation of the SM. However, part of the research in Physics beyond the SM consists in extending or relaxing the would-be fundamental symmetries that are manifest in the SM. In general, such a procedure allows for new physical effects that can potentially address open problems in fundamental Physics. As a particular example of such a program, Lorentz-symmetry breaking scenarios have been extensively studied over the past decades, see, e.g., \cite{Kostelecky:1988zi,Carroll:1989vb,Colladay:1996iz,Colladay:1998fq,Jackiw:1999yp,Kostelecky:2003fs,Kostelecky:2008ts,Diaz:2009qk,Belich:2009gaj,dePaulaNetto:2014kqi,dePaulaNetto:2017fpo,Mariz:2022oib}. Such investigations are mostly motivated by the possibility of taking into account new Physics (cumulatively) arising from very high-energy scales as, for example, from quantum-gravity models. There are many different ways to incorporate Lorentz-violating effects in the standard quantum-field theoretic setting but a particular framework in which this has been systematically investigated is the so-called Standard Model Extension (SME), see \cite{Colladay:1998fq}. The presence of a non-Abelian gauge sector in the SME serves as a motivation to investigate how the Lorentz-breaking terms may affect the standard properties of Lorentz-invariant YM theories. First studies on the renormalization of such a model at one-loop order were carried out in \cite{Colladay:2006rk} in linear covariant gauges and at lowest-order in the Lorentz-violating parameters. Subsequently, works on the renormalization of Lorentz-violating QCD and of the Electroweak sector in the presence of Lorentz-violating terms were carried out, see \cite{Colladay:2007aj,Colladay:2009rb}. In \cite{Santos:2014lfa}, the authors have carried out an algebraic proof of renormalizability at all orders in perturbation theory of Lorentz-violating YM (LVYM) theories in the Landau gauge. An intriguing outcome of such analysis is that, although the standard renormalization properties of YM theories are preserved in this scenario, a mass term for the non-Abelian gauge fields is generated. In other words, a mass-like term is necessary to make the theory perturbatively renormalizable according to \cite{Santos:2014lfa}. In \cite{Santos:2016uds} the authors have investigated the spectrum of such massive LVYM theory and found evidence for perturbative unitarity of the model. One important remark is in order: The algebraic analysis performed in \cite{Santos:2014lfa} employed what is sometimes referred to as Symanzik's method \cite{Symanzik:1969ek}, i.e., the target theory is embedded in a larger theory with more symmetries constructed upon the introduction of a suitable set of external sources. The target theory is regained by taking the physical values of such sources. In the context of LVYM, one can promote the fixed background tensors that are responsible to break Lorentz invariance to sources which in turn restore Lorentz symmetry. All terms that are compatible with the symmetries of the larger theory and with power counting will be generated and must be included in the starting point action. Consequently, after the algebraic characterization of the most general counterterm, the physical limit of such sources must be taken and in the case of LVYM, this entails the generation of a quadratic term on the gauge field with appropriate contractions of the fixed background tensors hence engendering a mass-like term. In the larger theory, those terms are compatible with BRST invariance thanks to the choice of introducing the underlying external sources in the form of BRST doublets. However, after the physical limit, BRST symmetry is softly broken by such mass terms. This may suggest that LVYM theories are not consistent. Nevertheless, the analysis carried out in \cite{Santos:2014lfa} was restricted to the Landau gauge which does not have any free gauge parameter. As such, one is not able to verify whether correlation functions of gauge-invariant operators are indeed gauge-parameter independent. In this work, we will show that, in fact, such a BRST breaking is harmless and a massive LVYM theory can be formulated in harmony with gauge invariance. The model that we will introduce reduces to the one obtained in \cite{Santos:2014lfa} in the Landau gauge. Although our aim is to introduce a LVYM massive model, we will also present some comments which suggest that the necessity of including the mass-like terms for the gauge fields is actually related to the Symanzik's method employed in \cite{Santos:2014lfa}. This means that if such a method was not used, our argument indicates that no mass terms would be generated. Hence, renormalizability of LVYM theories is ensured even in the absence of such mass-like terms. An explicit verification of this fact is relegated to future work.

In recent years, a massive extension of the gauge-fixed YM action in the Landau gauge has been scrutinized thanks to its ability to generate results at one- or two-loops for correlation functions that agree very well with gauge-fixed lattice simulations, see, e.g., \cite{Tissier:2010ts,Tissier:2011ey,Reinosa:2017qtf,Barrios:2024ixj,Gracey:2019xom,Barrios:2020ubx}. From this perspective, studying a massive extension of gauge-fixed YM theories can be seen as an effective way to access the non-perturbative behavior of pure YM theories by means of standard perturbative techniques. Hence, analyzing the structure of a massive extension of LVYM theories corresponds to an immediate generalization of such a program in the context of Lorentz-breaking theories. One potential first-principle justification for the inclusion of a mass-like term in the gauge-fixed action of YM theories is to account for the existence of Gribov copies. In \cite{Gribov:1977wm}, it was pointed out that the Landau-gauge condition does not fix the gauge redundancy completely. This means that field configurations which satisfy the gauge condition and are connected by a gauge transformation exist. This is not a particular problem of Landau gauge but rather a general obstruction in non-Abelian gauge theories. Therefore, the Faddeev-Popov (FP) gauge-fixing procedure \cite{Faddeev:1967fc} requires an improvement in order to remove those spurious configurations known as Gribov copies. This was worked out at leading order by Gribov in \cite{Gribov:1977wm} and extended to all orders by Zwanziger in \cite{Zwanziger:1989mf} and, in both cases, the removal of copies is restricted to those generated by infinitesimal gauge transformations. However, gauge copies associated with finite gauge transformations do exist as pointed out in \cite{vanBaal:1991zw}. In practice, the removal of infinitesimal Gribov copies in the Landau gauge is achieved by the restriction of the path integral to the so-called Gribov region which is free of infinitesimal Gribov copies (but still contains finite ones). The reader is referred to \cite{Sobreiro:2005ec,Vandersickel:2012tz} for reviews that explain the implementation of such a restriction in detail. Effectively, such a restriction generates a non-local modification of the gauge-fixed YM action in the Landau gauge which essentially introduces a mass-parameter to the gauge field propagator\footnote{The non-local modification is the so-called Horizon function. At leading order, such a modification looks like a momentum-dependent mass-like term for the gauge field.} which is not free but determined by a gap equation. The non-locality can be lifted by the introduction of auxiliary fields rendering the so-called Gribov-Zwanziger (GZ) action. The GZ action is a local and renormalizable way of removing infinitesimal Gribov copies from the YM path integral quantized in the Landau gauge \cite{Zwanziger:1989mf}. Due to the complicated structure of the GZ action involving extra fields, a simple massive extension of the gauge-fixed YM action provides an effective model to deal with infinitesimal Gribov copies.

Gribov copies are harmless in the perturbative regime of YM theories. Nevertheless, at strong coupling, the assumptions behind the FP procedure do not hold anymore and there is no compelling reason to not take into account the existence of gauge copies. Hence, Gribov copies can play an important role in the non-perturbative regime of YM theories. In fact, in \cite{Dudal:2007cw,Dudal:2008sp} it was shown that by including further non-perturbative effects in the GZ action, one is able to capture the non-perturbative behavior of the gauge-field propagator computed in gauge-fixed lattice simulations \cite{Cucchieri:2007rg,Duarte:2016iko,Dudal:2010tf,Cucchieri:2011ig,Dudal:2018cli} already at the tree-level of the modified GZ action. This new paradigm is known as the Refined Gribov-Zwanziger (RGZ) framework. More recently, it was shown that the tree-level properties are stable against one-loop corrections, see \cite{deBrito:2024ffa}. In \cite{Granado:2017xjs,Granado:2018bcp}, a systematic investigation of Gribov copies in LVYM was initiated. In the present work, we will formulate the RGZ action in the context of LVYM and its massive extension for a general linear covariant gauge. Such a construction will be performed in harmony with BRST symmetry. In summary, this work establishes a massive extension of the LVYM theories which is compatible with the theory described in \cite{Santos:2014lfa} and extends it to generic linear gauges without spoling BRST invariance. Furthermore, we argue that such a mass-like term is induced by the method employed in \cite{Santos:2014lfa} and is not to be expected in standard explicit perturbative computations. Yet the massive LVYM model corresponds to a potential effective descritpion to the infrared behavior of the theory encoding the presence of Gribov copies and the ensuing generation of condensates non-perturbatively.

This paper is organized as follows: In Sect.~\ref{Sec:Model}, we define the action of LVYM theories and its massive extension. Moreover, we present the appropriate gauge-fixing action derived from the BRST quantization of the underlying action. In Sect.~\ref{Sec:LandauGauge}, we comment on how the model just defined in the previous section reduced to the one introduced in \cite{Santos:2014lfa} in the Landau gauge. We also present an extensive discussion on the potential reasons behind the non-generation of a mass-like term for the gluons in pure LVYM and why Symanzik's method might not be suitable to be employed with the algebraic renormalization framework in this case. In Sect.~\ref{SecGCNPGF}, we discuss the Gribov problem in the context of LVYM. We start by a short overview of the problem in YM. Afterwards, we discuss how the removal of infinitesimal Gribov copies can be achieved in LVYM theories. Finally, we compute the tree-level gauge-field propagator which suffers major modifications thanks to the massive parameters that arise in the elimination of Gribov copies. A detailed analysis of the rather involved pole-structure of the propagator is not performed in this work and we concentrate on showing that the non-renormalizability of the standard linear covariant gauge remains intact in LVYM theories even after the elimination of infinitesimal Gribov copies. Conclusions and perspectives are finally presented.

%-------------------------------------------------------
\section{Definition of the Model \label{Sec:Model}}
%-------------------------------------------------------

%-------------------------------------------------------
\subsection{The classical action} \label{SubSec:ClassAction}
%-------------------------------------------------------

The action $S_{\rm LVYM}$ of LVYM theories in four Euclidean dimensions with  gauge group being $SU(N)$ is defined as
\begin{equation}
S_{\rm LVYM} = S_{\rm YM} + S_{\rm LVE} + S_{\rm LVO}\,,
\label{Eq:Model.1}
\end{equation}
with\footnote{We employ the short-hand notation $ \int {\rm d}^4 x \equiv  \int_x $.} 
\begin{equation}
S_{\rm YM} = \frac{1}{4} \int_x~F^{a}_{\mu\nu}F^{a}_{\mu\nu}\,,
\label{Eq:Model.2}
\end{equation}
where $F^a_{\mu\nu} = \partial_\mu A^a_\nu - \partial_\nu A^a_\mu + gf^{abc}A^b_{\mu} A^c_{\nu}$ and $g$ represents the gauge coupling. The symbol $f^{abc}$ denotes the structure constants of the gauge group $SU(N)$. The contribution $S_{\rm LVE}$ corresponds to the CPT-even contribution defined as\footnote{For convenience, we follow very closely the notation and conventions of \cite{Santos:2014lfa}.}
\begin{equation}
S_{\rm LVE} =  \int_x~\kappa_{\alpha\beta\mu\nu}F^{a}_{\alpha\beta}F^{a}_{\mu\nu}\,,
\label{Eq:Model.3}
\end{equation}
and $S_{\rm LVO}$ stands for the CPT-odd Lorentz-violating term, i.e.,
\begin{equation}
S_{\rm LVO} =  \int_x~v_\mu\epsilon_{\mu\nu\alpha\beta}\Bigg(A^a_\nu \partial_\alpha A^a_{\beta} + \frac{g}{3}f^{abc} A^a_{\nu}A^b_{\alpha}A^c_{\beta}\Bigg)\,.
\label{Eq:Model.4}
\end{equation}
The tensor $k_{\alpha\beta\mu\nu}$ and the vector $v_\mu$ are background structures that entail the violation of Lorentz invariance. The symbol $\epsilon_{\mu\nu\alpha\beta}$ is the totally antisymmetric Levi-Civita symbol.

The model we aim at defining corresponds to, on top of the action \eqref{Eq:Model.1}, add a gauge-invariant mass-like term. This term is defined as
\begin{eqnarray}
S_{M} &=& \frac{1}{2}\int_x~\mathbb{M}_{\mu\nu}(A^{h})^{a}_{\mu}(A^{h})^{a}_{\nu}+\int_x~\tau^a\partial_\mu (A^{h})^{a}_{\mu}\nonumber\\
&+&\int_x~\bar{\eta}^a\partial_\mu D^{ab}_\mu (A^h)\eta^b\,,
\label{Eq:Model.5}
\end{eqnarray}
with 
\begin{equation}
\mathbb{M}_{\mu\nu} = \delta_{\mu\nu}m^2 + a_{\mu}a_{\nu}\,.
\label{Eq:Model.6}
\end{equation}
The vector $a_\mu$ has mass dimension one and $m^2$ stands for a mass parameter. In principle, $a_\mu$ can be identified with $v_\mu$ but we leave the possibility of having different background structures open\footnote{At this stage, it is unclear if it is mandatory to enforce $a_\mu$ to be related to the LV background structures that are already present in the action \eqref{Eq:Model.1} This issue should be settled by a systematic analysis of the renormalizability properties of the action \eqref{Eq:Model.9} which is too complicated to be addressed in the present work. We leave this analysis for future work.}. The gauge-invariant operator $A^h$ corresponds to a dressed gauge-invariant field and its construction is summarized, e.g., in the appendix of \cite{Capri:2015ixa}. It is expressed as
\begin{equation}
(A^h)^a_\mu T^a = h^\dagger A^a T^a h + \frac{i}{g}h^\dagger \partial_\mu h\,.
\label{Eq:Model.7}
\end{equation}
The matrices $\left\{T^a\right\}$, with $a = 1,\ldots , N^2-1$ are the generators of the $SU(N)$ gauge group. Furthermore,
\begin{equation}
h = {\rm e}^{ig \xi^a T^a}\,,
\label{Eq:Model.8}
\end{equation}
with $\xi^a$ standing for a Stuckelberg-like field introduced to cast $A^h$ in a non-polynomial but local form. The fields $\tau^a$, $\bar{\eta}^a$ and $\eta^a$ are auxiliary fields that are necessary to localize the dressed field $A^h$, see \cite{Capri:2016ovw,Capri:2017bfd} and their integration results in a non-local gauge-invariant expression for $A^h$. The gauge invariance of the mass term is ensured by the definition of the Stueckelberg-like field $\xi^a$ and its gauge transformation. For a gauge transformation defined by $\mathcal{U}\in SU(N)$, one has
\begin{equation}
A^a_\mu T^a \rightarrow A^{\prime\,a}_\mu T^a = \mathcal{U}^\dagger A^a_\mu T^a \mathcal{U}+\frac{i}{g}\mathcal{U}^\dagger \partial_\mu \mathcal{U}\,,
\label{Eq:Model.8.1}
\end{equation}
and
\begin{equation}
h \rightarrow h^\prime = \mathcal{U}^\dagger h\, \quad {\rm and} \quad (h^\dagger) \rightarrow (h^\dagger)^\prime = h^\dagger \mathcal{U}
\label{Eq:Model.8.2}
\end{equation}
This implies
\begin{equation}
(A^h_\mu ) \rightarrow (A^h_\mu)^\prime = (A^h_\mu)\,,
\label{Eq:Model.8.3}
\end{equation}
establishing the gauge invariance of the dressed field $A^h_\mu$ and thus of the mass-term. Upon integration of the fields $(\tau,\bar{\eta},\eta)$ and, subsequently, of $\xi^a$, the dressed field $A^h_\mu$ becomes a non-local expression written as\footnote{The expression below is written in matrix form, i.e., the gauge field is expressed as $A_\mu = A^a_\mu T^a$.}
\begin{eqnarray}
&&A^{h,a}_\mu T^a \equiv A^h_\mu = A_\mu - \frac{\partial_\mu}{\partial^2}\partial_\alpha A_\alpha + ig \left[ A_\mu, \frac{1}{\partial^2}\partial_\alpha  A_\alpha \right] \nonumber\\
&+&  \frac{ig}{2}\left[\frac{1}{\partial^2}\partial_\alpha A_\alpha, \partial_\mu \frac{1}{\partial^2}\partial_\beta A_\beta \right] + ig \frac{\partial_\mu}{\partial^2}\left[\frac{\partial_\nu}{\partial^2}\partial_\alpha A_\alpha, A_\nu \right]\nonumber\\
&+&i\frac{g}{2}\frac{\partial_\mu}{\partial^2}\left[\frac{\partial_\alpha A_\alpha}{\partial^2},\partial_\beta A_\beta\right] + \EuScript{O}(A^3)\,.
\label{Eq:Model.8.4}
\end{eqnarray}
As said above, the details of the construction of such an expression can be found, e.g., in the appendix of \cite{Capri:2015ixa}, but we also refer the reader to \cite{Delbourgo:1986wz,Delbourgo:1987np,Zwanziger:1990tn,Lavelle:1995ty,Dragon:1996tk,Ruegg:2003ps} for more details and properties of the dressed field $A^h_\mu$. One important property to be emphasized is that, apart from the first term in the expansion \eqref{Eq:Model.8.4}, which is simply the gauge field $A_\mu$, all the other terms contain a divergence of the gauge field, i.e., $\partial_\mu A_\mu$. Thus, in the Landau gauge, this highly non-local expression collapses to the gauge field itself (on-shell) providing already a hint on the special character of the Landau gauge that we shall explore later on.

The massive model is defined as
\begin{equation}
S_{0} = S_{\rm LVYM} + S_M\,.
\label{Eq:Model.9}
\end{equation}
This action is ready to be quantized. As such, it requires the introduction of a gauge-fixing term. This is the topic of next subsection where the BRST quantization is employed.
%-------------------------------------------------------
\subsection{BRST Quantization} \label{SubSec:BRSTQuantization}
%-------------------------------------------------------

The BRST quantization can be used in order to gauge fix the action defined in \eqref{Eq:Model.9}. As such, we add to \eqref{Eq:Model.9} a BRST-exact term chosen to be
\begin{eqnarray}
S_{\rm FP} &=& \s \int_x~\bar{c}^a\Bigg(\partial_\mu A^a_\mu + \zeta_{\mu\nu}\partial_\mu A^a_\nu - \frac{\alpha}{2}b^a - \mu^2 \xi^a\Bigg)\nonumber\\
&=& \int_x~b^a\Bigg(\partial_\mu A^a_\mu + \zeta_{\mu\nu}\partial_\mu A^a_\nu - \frac{\alpha}{2}b^a - \mu^2 \xi^a\Bigg)\nonumber\\
&+&\int_x~\bar{c}^a\mathcal{M}^{ab}(\zeta,\mu^2)c^b\,.
\label{Eq:Model.10}
\end{eqnarray}
with
\begin{equation}
\mathcal{M}^{ab}(\zeta,\mu^2) = \partial_\mu D^{ab}_{\mu} + \zeta_{\mu\nu}\partial_\mu D^{ab}_\nu + \mu^2 g^{ab}(\xi)\,,
\label{Eq:Model.11}
\end{equation}
being the Faddeev-Popov operator and $(\bar{c},c)^a$ are the Faddeev-Popov ghosts. The BRST-operator $\s$ is nilpotent, i.e., $\s^2=0$ and its action on the elementary fields is defined by
\begin{align}
\s A^a_\mu &= - D^{ab}_\mu c^b\,,  &&\s c^a = \frac{g}{2}f^{abc}c^b c^c\,, \nonumber \\ 
\s \bar{c}^a &= b^a\,, &&\s b^a = 0\,, \nonumber \\ 
\s \tau^a &= 0\,, && \s\bar{\eta}^a = 0\,,\nonumber \\
\s \eta^a &= 0\,, && \s\xi^a = g^{ab}(\xi)c^b\,,
\label{Eq:Model.13}
\end{align}
with
\begin{equation}
g^{ab}(\xi) = -\delta^{ab} + \frac{g}{2}f^{abc}\xi^c - \frac{g^2}{12}f^{amr}f^{mbq}\xi^q\xi^r + \EuScript{O}(g^3)\,.
\label{Eq:Model.14}
\end{equation}
The gauge-fixing defined in eq.\eqref{Eq:Model.10} is a generalized linear covariant gauge which two gauge parameters, $\alpha$ and $\mu^2$ and a tensor $\zeta_{\mu\nu}$, which is a generalization of gauge parameter, being, actually, a gauge matrix. To the best of our knowledge, there is no systematic analysis of the renomalization properties of such a gauge condition, especially in the context of LVYM theories (and its massive extension), and this is relegated to future work. For the present purposes, we emphasize the linear nature of this gauge choice and that observables should not depend on the choice of $(\alpha,\mu^2)$ and $\zeta_{\mu\nu}$. The particular case where $\alpha = \mu^2 = 0$ and $\zeta_{\mu\nu}= 0$ corresponds to the Landau gauge. Its renormalization properties were investigated in Lorentz violating Yang-Mills theories in \cite{Santos:2014lfa} by means of the algebraic renormalization framework \cite{Piguet:1995er}.

Therefore, the gauge-fixed LVYM action extended by a mass-like term is denoted by $S$ and defined as
\begin{equation}
S = S_{0} + S_{\rm FP}\,.
\label{Eq:Model.15}
\end{equation}
The action $S$ is invariant under the BRST transformations \eqref{Eq:Model.13}, i.e., $\s S =0$. As usual, one can define an extended action $\Sigma$ by coupling the non-linear BRST transformations to external sources. This action is defined as
\begin{equation}
\Sigma = S + \int_x~\Big(\Omega^a_\mu \s A^a_\mu + L^a \s c^a + K^a \s \xi^a\Big)\,.
\label{Eq:Model.16}
\end{equation}
The external sources $(\Omega^a_\mu,L^a,K^a)$ are BRST invariant by definition. It follows that $\s \Sigma = 0$. BRST invariance can also be phrased in terms of the Slavnov-Taylor identity for $\Sigma$ which reads
\begin{equation}
\EuScript{S}(\Sigma) = 0\,,
\label{Eq:Model17}
\end{equation}
with
\begin{eqnarray}
\EuScript{S}(\Sigma) = \int_x \Bigg(\frac{\delta \Sigma}{\delta A^a_\mu}\frac{\delta \Sigma}{\delta \Omega^a_\mu} + \frac{\delta \Sigma}{\delta c^a}\frac{\delta \Sigma}{\delta L^a} + \frac{\delta \Sigma}{\delta \xi^a}\frac{\delta \Sigma}{\delta K^a}+ b^a \frac{\delta \Sigma}{\delta \bar{c}^a}\Bigg)\,,\nonumber\\
\label{Eq:Model18}
\end{eqnarray}
which can be extended to the quantum action $\Gamma$, i.e., $\EuScript{S}(\Gamma)=0$. At this stage, a comment is in order: The tensor $\zeta_{\mu\nu}$ that appears in the gauge-fixing condition breaks Lorentz invariance in the sense of being a tensorial background structure. Yet this breaking is introduced in a BRST-exact term and therefore does not contribute to correlation functions of observables. In order words: this gauge-fixing is a perfectly valid choice even in a Lorentz-symmetry preserving scenario since the breaking induced by $\zeta_{\mu\nu}$ is completely unphysical. Nevertheless, for explicit calculations in Lorentz-violating extensions of YM theories, such a gauge-fixing condition can be advantageous since it can be used to simplify the form of the would-be gluon (gauge field) propagator by a suitable choice of $\zeta_{\mu\nu}$, see, e.g., \cite{Altschul:2023vnf}. This is different from the tensor $\mathbb{M}_{\mu\nu}$ (as well as the usual tensorial background structures in LVYM such as $\kappa_{\alpha\beta\mu\nu}$ and $v_\mu$) which arises coupled to a BRST-closed operator. In this case, there is no screening mechanism that forbids the presence of the structures of $\mathbb{M}_{\mu\nu}$ (as well as of $\kappa_{\alpha\beta\mu\nu}$ and $v_\mu$) in correlation functions of gauge-invariant operators. 

%-------------------------------------------------------
\section{The special case of Landau Gauge \label{Sec:LandauGauge}}
%-------------------------------------------------------

The aim of this section is to establish a connection between the model presented in the previous section and the LVYM model investigated in \cite{Santos:2014lfa}. In \cite{Santos:2014lfa}, the proof of renormalizability of the LVYM action was achieved in the Landau gauge by means of the algebraic renormalization technique \cite{Piguet:1995er}. The authors adopted the so-called Symanzik approach where the target theory is embedded in a theory which enjoys more symmetries thanks to the introduction of sources that transform suitably \cite{Symanzik:1969ek}. Lorentz invariance is recovered, in a certain sense, by the introduction of external sources which attain physical values at the end of the renormalization analysis. By this procedure, the authors verified that a mass term for the gauge field must be included for consistency. In simple words: the procedure adopted by the authors requires the introduction of a mass term for the gauge field in the Landau gauge. This finding raises a natural question: Can BRST symmetry be made compatible with such a massive extension of the LVYM theory? As we shall discuss, the Landau gauge is rather special and allows for a change of variables that renders the theory BRST invariant. On the other hand, it becomes rather obscure what would happen if one proceeds along the same way in a gauge different from the Landau choice. In particular, if such a mass-term is generated in linear covariant gauges, BRST symmetry would be broken and there is no change of variable that produces a BRST-invariant reformulation of the model. However, should the mass term be absent in linear covariant gauges, an immediate problem arises: the Landau gauge is a particular case of linear covariant gauge and thus, by adjusting the gauge parameter, a mass term should be generated according to the analysis of \cite{Santos:2014lfa}. This produces a clear problem: Either LVYM theories do require a mass term for their renormalizability and this would be at odds in different gauges than the Landau gauge leading to an inconsistent model or such a mass term is not generated in the Landau gauge as pointed out in \cite{Santos:2014lfa}. The model here proposed supplements the LVYM action with a gauge-invariant mass term and therefore can be viewed as a consistent generalization of the model introduced in \cite{Santos:2014lfa} for different gauges. Yet it is desirable to establish a precise relation between the model proposed in this work and the result presented in \cite{Santos:2014lfa}.

We start with the observation that $A^h_\mu$ can be expressed as\footnote{See the  appendix of \cite{Capri:2015ixa} for details.}
\begin{equation}
A^{h,a}_\mu = A^a_\mu - \frac{\partial_\mu}{\partial^2}(\partial \cdot A^a) + \mathcal{R}^{ab}_\mu (A) (\partial \cdot A^b)\,.
\label{Eq:LandauGauge.1}
\end{equation} 
After integrating out the auxiliary fields $(\tau,\xi,\bar{\eta},\eta)^a$, the mass term is written as\footnote{See, e.g., the Appendix of \cite{Capri:2017bfd} for a detailed discussion on the integration of those auxiliary fields.}. Hence,
\begin{equation}
A^{h,a}_\mu A^{h,a}_\nu = A^a_\mu A^a_\nu + \EuScript{F}^{a}_{\mu\nu} (A) (\partial \cdot A^a)\,,
\label{Eq:LandauGauge.2}
\end{equation}
The key property of eq.\eqref{Eq:LandauGauge.2} is that the ``dressed" quadratic term of $A^h$ term can be expressed as a quadratic term on the gauge field itself $A_\mu$ plus terms that necessarily contains a divergence of the gauge field. In the Landau gauge, the $\EuScript{F}(A)$ terms can be absorbed in a field redefinition of the $b$-field, i.e.,
\begin{equation}
b^a\,\,\to\,\,b^{h,a} = b^a+\frac{\mathbb{M}_{\mu\nu}}{2}\EuScript{F}^{a}_{\mu\nu} (A)\,.
\label{Eq:LandauGauge.3}
\end{equation} 
This field redefinition has trivial Jacobian and the underlying ``new" action is
\begin{eqnarray}
\tilde{S} &=& S_{\rm LVYM} + \int_x \Big(b^{h,a}\partial_\mu A^a_\mu +\bar{c}^a\partial_\mu D^{ab}_{\mu}c^b\Big)\nonumber\\
&+&\frac{\mathbb{M}_{\mu\nu}}{2}\int_x A^a_\mu A^a_\nu\,
\label{Eq:LandauGauge.4}
\end{eqnarray}
which is exactly the action proposed in \cite{Santos:2014lfa}. This establishes a clear relation between the BRST-invariant action \eqref{Eq:Model.15} and the soft BRST-broken action (by the mass term) \eqref{Eq:LandauGauge.4}. Of course, this relation is a very particular feature of the Landau gauge (notice that in a different gauge, the terms in $\EuScript{F}(A)$ would not be absorbed by a field redefinition of the $b$-field. As a consequence, correlation functions involving the gauge field and Faddeev-Popov ghosts are exactly the same in both theories, i.e.,
\begin{equation}
\langle \Phi^{A_1}(x_1)
\ldots \Phi^{A_n}(x_n)\rangle^L_{S} = \langle \Phi^{A_1}(x_1)
\ldots \Phi^{A_n}(x_n)\rangle^L_{\tilde{S}}\,,
\label{Eq:LandauGauge.5}
\end{equation}
with $\Phi^A (x) \in \left\{A^a_\mu (x), \bar{c}^a (x), c^a (x)\right\}$ and the superscript $L$ stands for Landau. In particular, correlation functions of gauge-invariant (BRST-closed) operators are the same in both formulations, even if $\tilde{S}$ breaks BRST-symmetry softly. This allows one to write the following identity
\begin{eqnarray}
\langle \EuScript{O}(x_1)
\ldots \EuScript{O}(x_n)\rangle^{\rm \alpha,\zeta,\mu^2}_{S} &=& \langle \EuScript{O}(x_1)
\ldots \EuScript{O}(x_n)\rangle^L_{S}\nonumber\\
&=& \langle \EuScript{O}(x_1)
\ldots \EuScript{O}(x_n)\rangle^L_{\tilde{S}}\,.
\label{Eq:LandauGauge.6}
\end{eqnarray}
Consequently, for the computation of observables in the massive extension of LVYM, using the action $\tilde{S}$ is sufficient (and simpler) for all practical purposes. Therefore, in the Landau gauge, working with the BRST-broken version of the action, i.e., $\tilde{S}$, is a perfectly valid choice for the computation of correlation functions of gauge-invariant operators or even of gauge fields and Faddeev-Popov ghosts. 

At this stage, it is relevant to make a brief summary and a couple of remarks: The model introduced in the present paper is a massive extension of LVYM that is compatible with gauge/BRST-invariance. By taking the particular choice of the Landau gauge, the action of such a model reduces to the action introduced in \cite{Santos:2014lfa} which had massive terms arising due to the application of Symanzik's method together with the algebraic renormalization procedure. The model introduced in \cite{Santos:2014lfa} breaks BRST-invariance explicitly but we just showed in the present section that it can be made BRST-invariant by a suitable field redefinition. By letting the gauge parameters free one can analyze the renormalization properties of action \eqref{Eq:Model.15}. This is left for future work. This brings us to a delicate situation: If one ignores the mass-like term in the LVYM extended action and perform its algebraic renormalization analysis using the prescriptions employed in \cite{Santos:2014lfa} in a linear covariant gauge, either a mass-term is generated or it is not. If it is not generated, this is problematic because linear covariant gauges should deform to the Landau gauge within suitable choices of gauge parameters and this would contradict the result in \cite{Santos:2014lfa}. However, if the (naive) mass term is generated, it certainly breaks BRST and cannot be removed by the field-redefinition discussed in the present section. In this sense, those models would be completely inconsistent with gauge/BRST invariance. To our understanding, this opens the following possibilities: 1) There is no mass generation and the results of \cite{Santos:2014lfa} are due to the method employed or 2) LVYM cannot be consistent with BRST invariance and hence will generate gauge-dependent correlation functions for would-be gauge-invariant operators, i.e., LVYM would be inconsistent. A third possibility is to define LVYM theories already with the gauge-invariant mass term as proposed in this paper. This would ensure gauge invariance, although many issues are yet to be explored in this model. Our preliminary conclusion that requires further investigations to be substantiated is that the mass generation seen in \cite{Santos:2014lfa} is a consequence of the application of Symanzik's method. The reason is the following: Symanzik's method \cite{Symanzik:1969ek} is designed for softly broken symmetries, i.e., Lagrangians or actions that have an underlying explicit symmetry breaking thanks to the presence of a term that is quadratic on the fields. In this sense, the symmetry-breaking term does not generate new vertices and if the symmetric theory is renormalizable so is the symmetry-broken theory. This technique is employed, e.g., in the context of the GZ
 action as originally formulated due to the presence of a soft BRST-symmetry breaking term, see, e.g., \cite{Vandersickel:2012tz}. In \cite{Santos:2014lfa}, the Lorentz violating terms \eqref{Eq:Model.3} and \eqref{Eq:Model.4} are not quadratic on the the fields. Hence, to our knowledge, the application of Symanzik's prescription does not ensure that the symmetric theory after the introduction of symmetry-restoring sources will be the same as the symmetry-broken one. As such, we envision that an algebraic renormalization analysis of the LVYM theories without Symanzik's method will not produce a mass term for the gauge field. Such a mass term found in \cite{Santos:2014lfa} is proportional to the background tensors that break Lorentz invariance and its generation is tied to the physical values attained to the symmetry-restoring sources. The algebraic renormalization of LVYM without Symanzik's prescription was not performed even in the Landau gauge, to our knowledge and it would be a very important aspect to be understood in this class of theories. There are another two important issues to be emphasized in this regard: Also in \cite{Santos:2014lfa}, the same procedure is applied to the Abelian version of LVYM. In this case, no mass generation is seen. However, the Lorentz-violating terms are all quadratic on the (Abelian) gauge field. Hence, Symanzik's procedure should give the same renormalization properties for both symmetric and symmetry-broken theories. This is consistent with our explanation above. The second issue to be stressed is that, ultimately, one can always perform an explicit computation and renormalize LVYM theories at, e.g., one-loop order. We stress two examples of works where this was done, i.e., \cite{Colladay:2006rk,Altschul:2023vnf}. In \cite{Colladay:2006rk}, the analysis was performed for pure Yang-Mills as discussed in the present paper while in \cite{Altschul:2023vnf}, the one-loop computation included also the matter coupling with scalar degrees of freedom. In both papers, no signal of mass generation of the gauge field seems to appear. Nevertheless, in both cases the computations are limited to the lowest order in Lorentz-violating parameters. The mass term generated in the anslysis of \cite{Santos:2014lfa} comes along with quadratic or higher-order Lorentz-violating terms. Hence, the explicit loop computations that are available in the literature cannot be conclusive in terms of the generation or not of mass terms for the gauge fields. For that, one would need to go at least one order higher in the expansion of the Lorentz-violating structures. This computation will be reported somewhere else. To close this Section, we remark that the generation of a mass term for gauge fields in LVYM is still an open issue and we have raised some arguments that possibly support the non-existence of such a term. Nevertheless, one of the main points of this work is to argue that one can still define a massive LVYM theory that is compatible with gauge/BRST invariance as a stand alone model that can be useful to explore non-perturbative properties of LVYM theories since they emulate effects arising from the existence of Gribov copies. In the next section, we discuss the removal of infinitesimal Gribov copies in LVYM and its consequences on the tree-level propagator of the gauge field. Finally, we point out that the massive-model introduced in \cite{Santos:2014lfa} features a BRST-like symmetry in the case of $a_\mu \to 0$, however, nilpotent. This can be achivied by the following transformations:
\begin{align}
\s A^a_\mu &= - D^{ab}_\mu c^b\,,  &&\s c^a = \frac{g}{2}f^{abc}c^b c^c\,, \nonumber \\ 
\s \bar{c}^a &= b^a\,, &&\s b^a = -m^2 c^a\,.
\label{Eq:Model.Deformed.BRST}
\end{align}
Such a deformation of the BRST symmetry is simply a formal observation since the BRST transformations are not nilpotent and this plays a pivotal role in the use of BRST-invariance. Moreover, it requires that the mass-like term does not possess the background structure introduced in this work as $a_\mu$. We are not aware if a generalization that incorporate $a_\mu$ is viable but, in any case, such a non-nilpotent invariance has very limited application.

%-------------------------------------------------------
\section{Gribov Copies and Non-perturbative Gauge-fixing \label{SecGCNPGF}}
%-------------------------------------------------------

%-------------------------------------------------------
\subsection{Short overview of the Gribov problem in Yang-Mills theories \label{SubSecGPYM}}
%-------------------------------------------------------

This section is devoted to the discussion of the existence of infinitesimal Gribov copies in the Landau gauge (and its generalization to more generic gauges) and the effects they generate to LVYM and its massive extension. Irrespective of the underlying dynamics, the Landau gauge is plagued by (infinitesimal) Gribov copies due to the existence of normalizable zero modes of the Faddeev-Popov operator \cite{Gribov:1977wm,Guimaraes:2011sf,Capri:2012ev}. If a gauge field configuration $A^a_\mu$ satisfies the Landau gauge condition, i.e., $\partial_\mu A^a_\mu = 0$ then a gauge equivalent configuration $A^{\prime\,a}_{\mu}$ satisfies the Landau gauge condition as well if
\begin{equation}
\partial_\mu A^{\prime\, a}_\mu = 0 \,\, \Rightarrow \,\, \partial_\mu (A^a_\mu - D^{ab}_{\mu}\theta^b) = 0 \,\, \Rightarrow \,\, -\partial_\mu D^{ab}_{\mu}\theta^b = 0\,,
\label{Eq:GCNPGF.1}
\end{equation} 
with $\theta^a$ being an infinitesimal parameter of the gauge transformation. Hence, field configurations that satisfy the Landau gauge condition and are related to other field configurations which satisfy the Landau gauge condition by an infinitesimal gauge transformation, must generate a (normalizable) zero-mode of the FP operator. As pointed out by Gribov in \cite{Gribov:1977wm}, the FP operator has normalizable zero modes and therefore those configurations exist. This corresponds to a residual gauge redundancy and those spurious configurations are known as (infinitesimal) Gribov copies. This violates one of the assumptions of the FP procedure which takes for granted that the gauge conditions select one representative per gauge orbit. Notice that the discussion at this point is restricted to infinitesimal Gribov copies, but they are not the whole story. In fact, there are Gribov copies generated by finite gauge transformations as discussed in \cite{vanBaal:1991zw}. In fact, the existence of Gribov copies is not a particular issue of the Landau gauge but a manifestation of the non-trivial geometrical structure of non-Abelian gauge theories as discussed in \cite{Singer:1978dk}.

The elimination of infinitesimal Gribov copies was proposed by Gribov himself in \cite{Gribov:1977wm} in pure YM theories at leading order in perturbation theory. In \cite{Zwanziger:1989mf}, Zwanziger proposed an alternative prescription which was generalized to all orders. An equivalence of both approaches was is established in \cite{Capri:2012wx}. The core of the procedure responsible for the elimination of infinitesimal Gribov copies is the so-called Gribov region $\Omega$. It is defined as
\begin{equation}
\Omega = \left\{A^a_\mu\,, \partial_\mu A^a_\mu =0 \,\Big|\, -\partial_\mu D^{ab}_{\mu} > 0\right\}\,.
\label{Eq:GCNPGF.2}
\end{equation} 
Such a definition is possible because the FP operator in the Landau gauge is Hermitian, allowing for a meaningful positivity statement regarding its spectrum. The Gribov region satisfies a number of important properties. In particular, it is bounded in every direction in field space (and its boundary is known as the Gribov horizon), it is convex, and every gauge orbits cross it at least once, see \cite{DellAntonio:1991mms}. Effectively, the elimination of infinitesimal Gribov copies in the Landau gauge consists in restricting the path integral measure to $\Omega$,
\begin{equation}
\EuScript{Z} = \int_\Omega [\EuScript{D}\mu]_{\rm YM}\,{\rm e}^{-S_{\rm YM}-S_{\rm FP}}\,.
\label{Eq:GCNPGF.3}
\end{equation}
The imposition of the restriction of the path integral measure to $\Omega$ is a purely geometrical procedure and does not depend on the particular form of the gauge-invariant action (in eq.\eqref{Eq:GCNPGF.3}, we have chosen the YM action for concreteness, but it could be a different choice such as the LVYM action). The path integral measure in eq.\eqref{Eq:GCNPGF.3} is defined as
\begin{equation}
[\EuScript{D}\mu]_{\rm YM} = [\EuScript{D}A][\EuScript{D}b][\EuScript{D}\bar{c}][\EuScript{D}c]\,.
\label{Eq:GCNPGF.4}
\end{equation}
Effectively, the restriction to $\Omega$ is imposed by a modification of the Boltzmann weight in \eqref{Eq:GCNPGF.3} as follows,
\begin{equation}
\EuScript{Z} = \int [\EuScript{D}\mu]_{\rm YM}\,{\rm e}^{-S_{\rm YM}-S_{\rm FP}+\gamma^4 H(A) - 4V\gamma^4 (N^2-1)}\,.
\label{Eq:GCNPGF.5}
\end{equation}
The function $H(A)$ is the so-called horizon function and has the explicit non-local form
\begin{equation}
H(A) = g^2 \int_{x,y} f^{abc}A^b_\mu (x) \Big[\EuScript{M}^{-1}\Big]^{ad} (x,y) f^{dec} A^e_{\mu}(y)\,,
\label{Eq:GCNPGF.6}
\end{equation}
with $\EuScript{M}^{ab} = \mathcal{M}^{ab}(0,0) \equiv -\partial_\mu D^{ab}_\mu$ being the FP operator in the Landau gauge. The total volume of the four-dimensional space(time) is represented by $V$. The parameter $\gamma^2$ is known as the Gribov parameter. It is not free, but fixed self-consistently by a gap equation
\begin{equation}
\langle H(A) \rangle = 4V (N^2-1)\,.
\label{Eq:GCNPGF.7}
\end{equation}
The non-local form of the horizon function can be evaded by the introduction of a suitable set of auxiliary fields $(\bar{\varphi},\varphi)^{ab}_{\mu}$ and $(\bar{\omega},\omega)^{ab}_{\mu}$ of bosonic and Grassmannian nature, respectively. The localized action and its partition function are given below,
\begin{equation}
\EuScript{Z} = \int [\EuScript{D}\mu]_{\rm GZ}\,{\rm e}^{-S_{\rm GZ}- 4V\gamma^4 (N^2-1)}\,,
\label{Eq:GCNPGF.8}
\end{equation}
with
\begin{eqnarray}
S_{\rm GZ} &=& S_{\rm YM} + S_{\rm FP} - \int_x \Big(\bar{\varphi}^{ac}_{\mu}\EuScript{M}^{ab}\varphi^{bc}_\mu - \bar{\omega}^{ac}_{\mu}\EuScript{M}^{ab}\omega^{bc}_\mu\Big)\nonumber\\
&-&\gamma^2 \int_x gf^{abc}A^a_\mu (\varphi + \bar{\varphi})^{bc}_\mu\,.
\label{Eq:GCNPGF.9}
\end{eqnarray}
The action \eqref{Eq:GCNPGF.9} is known as the Gribov-Zwanziger (GZ) action. It is local and renormalizable to all orders in perturbation theory \cite{Zwanziger:1989mf}. The functional measure is given by
\begin{equation}
[\EuScript{D}\mu]_{\rm GZ} = [\EuScript{D}\mu]_{\rm YM}[\EuScript{D}\bar{\varphi}][\EuScript{D}{\varphi}][\EuScript{D}\bar{\omega}][\EuScript{D}{\omega}]\,.
\label{Eq:GCNPGF.10}
\end{equation}
In the local form, the gap equation that fixes the Gribov parameter $\gamma^2$ is written as
\begin{equation}
\frac{\partial \EuScript{E}_v}{\partial \gamma^2} = 0\,,
\label{Eq:GCNPGF.11}
\end{equation}
with $\EuScript{E}_v$ being the vacuum energy. The GZ action features many interesting properties that we will not discuss in the present work and we refer the interested reader to, e.g., \cite{Vandersickel:2012tz}. However, a particular feature of the action \eqref{Eq:GCNPGF.9} is that it breaks BRST symmetry explicitly. Yet the breaking is soft in the sense that it vanishes in the deep ultraviolet where the standard gauge-fixed Yang-Mills action is recovered. The breaking of BRST symmetry in the GZ framework was deeply investigated over the past years, see, e.g.,  \cite{Maggiore:1993wq,Baulieu:2008fy,Dudal:2009xh,Sorella:2009vt,Sorella:2010it,Capri:2010hb,Lavrov:2011wb,Serreau:2012cg,Serreau:2013ila,Dudal:2012sb,Pereira:2013aza,Pereira:2014apa,Lavrov:2013boa,Capri:2014bsa,Cucchieri:2014via,Moshin:2014xka,Schaden:2014bea,Schaden:2015uua}. However, in \cite{Capri:2015ixa}, a reformulation of the GZ action which enjoys an exact and nilpotent BRST invariance was proposed. The key step in this reformulation was the use of the dressed gauge field $A^h_\mu$ already presented in Eq.~\eqref{Eq:Model.7}. In this reformulation, the BRST-invariant GZ action is written as
\begin{eqnarray}
S^{\rm inv}_{\rm GZ} &=& S_{\rm YM} + S_{\rm FP} - \int_x \Big(\bar{\varphi}^{ac}_{\mu}\EuScript{M}^{ab}(A^h)\varphi^{bc}_\mu \nonumber\\ 
&-& \bar{\omega}^{ac}_{\mu}\EuScript{M}^{ab}(A^h)\omega^{bc}_\mu\Big)
-\gamma^2 \int_x gf^{abc}(A^h)^a_\mu (\varphi + \bar{\varphi})^{bc}_\mu\,\nonumber\\
&+&\int_x~\tau^a\partial_\mu (A^{h})^{a}_{\mu}
+\int_x~\bar{\eta}^a\partial_\mu D^{ab}_\mu (A^h)\eta^b\,.
\label{Eq:GCNPGF.12}\nonumber\\
\end{eqnarray}
All auxiliary fields transform as BRST singlets. In practice, it corresponds to the replacement of $A_\mu$ by $A^h_\mu$ in those terms arising from the restriction of the path integral to the Gribov region. In eq.~\eqref{Eq:GCNPGF.12}, the last two terms are the same introduced in \eqref{Eq:Model.5} to localize the field\footnote{The BRST-invariant formulation of the GZ action has two sorts of non-localities. The integration of the localizing fields generate the non-local structure of the horizon function as well as the non-local version of $A^h_\mu$.} $A^h_\mu$.

It was realized in \cite{Dudal:2007cw,Dudal:2008sp} that the GZ action suffers from infrared instabilities and favors the formation of condensates. In particular, the localizing fields acquire their own dynamics and condense in the infrared. It was proposed a refinement of the GZ action leading to the Refined GZ (RGZ) action. In its BRST-invariant formulation, the action is expressed as
\begin{eqnarray}
S^{\rm inv}_{\rm RGZ} &=& S^{\rm inv}_{\rm GZ} + \frac{m^2}{2}\int_x (A^h)^a_\mu(A^h)^a_\mu - M^2\int_x \Big(\bar{\varphi}^{ab}_{\mu}\varphi^{ab}_\mu\nonumber\\
&-&\bar{\omega}^{ab}_{\mu}\omega^{ab}_\mu\Big)\,.
\label{Eq:GCNPGF.13}
\end{eqnarray}
This BRST-invariant formulation of the RGZ action in the Landau gauge is related to the standard BRST-broken formulation \cite{Dudal:2007cw,Dudal:2008sp} by a suitable change of variables. The tree-level gluon propagator obtained from \eqref{Eq:GCNPGF.13} attains a finite value at vanishing momentum and fits well data extracted from very large lattices, see \cite{Cucchieri:2007md,Cucchieri:2007rg,Sternbeck:2007ug} for this correlation function. More recently, the gluon propagator was computed at one-loop order within the RGZ scenario and the results make evident that the one-loop corrections do not destroy the important features of the tree-level gluon propagator, see \cite{deBrito:2024ffa}. Studies of the ghost-gluon vertex and scalar-matter coupling to \eqref{Eq:GCNPGF.13} are in good agreement with lattice simulations as well, see \cite{Mintz:2017qri,deBrito:2023qfs,Barrios:2024idr} and references therein. It is important to emphasize that the Gribov problem and the proposal of effective actions,  such as the RGZ action, that remove infinitesimal copies is possible in a BRST-invariant setup. This was established in several works \cite{Capri:2015ixa,Capri:2015nzw,Capri:2016aqq,Capri:2016gut,Pereira:2016fpn,Capri:2017abz} and a proof of renormalizability in linear covariant gauges was achieved in \cite{Capri:2017bfd}. We remark once again that the (R)GZ framework deals with the removal of infinitesimal Gribov copies. In principle, the resulting action still does not account for finite gauge copies and up to date there is not systematic solution to this problem. In the following, we discuss the impact of the restriction of the path integral measure to the Gribov region to the infrared dynamics of LVYM theories.
%-------------------------------------------------------
\subsection{Consequences of the removal of infinitesimal Gribov copies in LVYM theories \label{SubSecGPLVYM}}
%-------------------------------------------------------

The gauge-fixed LVYM action in the Landau gauge displays the same residual gauge invariance due to the existence of Gribov copies. Due to the geometrical nature of the problem, the equation which defines the infinitesimal Gribov copies does not depend on the choice of the gauge-invariant action. Moreover, the restriction of the path integral measure to the Gribov region is achieved by the same procedure outlined in the previous subsection. As such, the partition function for LVYM theories quantized in the Landau gauge and restricted to the Gribov region reads
\begin{equation}
\EuScript{Z} = \int [\EuScript{D}\mu]_{\rm GZ}\,{\rm e}^{-S_{\rm LVYM}-S_{\rm FP}+\gamma^4 H(A) - 4V\gamma^4 (N^2-1)}\,.
\label{Eq:GPLVYM.1}
\end{equation}
By means of the localizing procedure discussed above, we define the local version of the Griboz-Zwanziger framework applied to LVYM theories in the Landau gauge, i.e., 
\begin{eqnarray}
S^{\rm LV}_{\rm GZ} &=& S_{\rm LVYM} + S_{\rm FP} - \int_x \Big(\bar{\varphi}^{ac}_{\mu}\EuScript{M}^{ab}\varphi^{bc}_\mu \nonumber\\ 
&-& \bar{\omega}^{ac}_{\mu}\EuScript{M}^{ab}\omega^{bc}_\mu\Big)
-\gamma^2 \int_x gf^{abc}A^a_\mu (\varphi + \bar{\varphi})^{bc}_\mu\,.\label{Eq:GPLVYM.2}
\end{eqnarray}
This action breaks BRST symmetry but it can be extended to a BRST-invariant formulation by means of the introduction of the dressed gauge field $A^h_\mu$ in a completely equivalent fashion as discussed in the previous subsection. It reads,
\begin{eqnarray}
S^{\rm inv}_{\rm LVGZ} &=& S_{\rm LVYM} + S_{\rm FP} - \int_x \Big(\bar{\varphi}^{ac}_{\mu}\EuScript{M}^{ab}(A^h)\varphi^{bc}_\mu \nonumber\\ 
&-& \bar{\omega}^{ac}_{\mu}\EuScript{M}^{ab}(A^h)\omega^{bc}_\mu\Big)
-\gamma^2 \int_x gf^{abc}(A^h)^a_\mu (\varphi + \bar{\varphi})^{bc}_\mu\nonumber\\
&+&\int_x~\tau^a\partial_\mu (A^{h})^{a}_{\mu}
+\int_x~\bar{\eta}^a\partial_\mu D^{ab}_\mu (A^h)\eta^b\,.
\label{Eq:GPLVYM.3}
\end{eqnarray}
Next to that, we add the refining terms. Here, one comment is in order: Thanks to the structure of the gauge-invariant mass term introduced in \eqref{Eq:Model.5}, we will add a mass-like term for the gauge field which accommodates a non-trivial tensorial structure\footnote{In fact, it is necessary to explicitly check if the non-trivial tensorial structure is generated due to infrared instabilities or not. For the time being, we add in order to contemplate the situation in which such a mass term is introduced from the beginning in a gauge-invariant fashion.}. In principle, the localizing fields could also have such a non-trivial tensorial structure. In order to find out if those terms are generated, a detailed analysis of those composite operators should be performed. In fact, a careful analysis of the renormalizability properties of \eqref{Eq:GPLVYM.3} is necessary. This falls beyond the scope of the present paper and for this reason, we just introduce the standard Lorentz-invariant refining condensates in the auxiliary-fields sector. Hence, the RGZ action arising from the LVYM theory in the Landau gauge is given by
\begin{eqnarray}
S^{\rm inv}_{\rm LVRGZ} &=& S^{\rm inv}_{\rm LVGZ} + \frac{\mathbb{M}_{\mu\nu}}{2}\int_x (A^h)^a_\mu(A^h)^a_\nu \nonumber\\
&-& M^2\int_x \Big(\bar{\varphi}^{ab}_{\mu}\varphi^{ab}_\mu
-\bar{\omega}^{ab}_{\mu}\omega^{ab}_\mu\Big)\,.
\label{Eq:GPLVYM.4}
\end{eqnarray}
This action is local, effectively eliminates infinitesimal Gribov copies from the path integral of LVYM theories quantized in the generic gauges defined in \eqref{Eq:Model.10} and takes into account the effects of dimension-two condensates in a BRST-invariant fashion. Clearly, due to the non-polynomial nature of $(A^h)^a_\mu$, the action is non-polynomial on the local fields. However, within the Landau gauge $(\zeta_{\mu\nu}=\alpha=\mu^2=0)$, the non-polynomial nature of this action can be circumvented thanks to the decoupling of the Stueckelberg-like field $\xi^a$. Following the same reasoning discussed in Sect.~\ref{Sec:LandauGauge}, one is led to
\begin{eqnarray}
\langle \Phi^{A_1}(x_1)
&\ldots &  \Phi^{A_n}(x_n)\rangle^{\rm Landau}_{S^{\rm inv}_{\rm LVRGZ}} \nonumber\\
 &=& \langle \Phi^{A_1}(x_1)
\ldots \Phi^{A_n}(x_n)\rangle_{S^{\rm LV}_{\rm RGZ}}\,.
\label{Eq:GPLVYM.5}
\end{eqnarray}
For all practical purposes, using the action $S^{\rm LV}_{\rm RGZ}$ to perform explicit computations is way simpler. Yet the BRST invariance of $S^{\rm inv}_{\rm LVRGZ}$ makes it clear that the newly introduced mass parameters $(\gamma^2,M^2)$ together with the ones introduced in $\mathbb{M}_{\mu\nu}$ are not akin to gauge parameters since they are coupled to BRST-closed operators.

%-------------------------------------------------------
\subsection{Gauge field propagator at tree-level \label{SubSecTreeLevelGFP}}
%-------------------------------------------------------

The gauge-field propagator obtained at tree-level in the standard Lorentz-symmetric RGZ framework can be fitted quite well with lattice simulations in the deep infrared, see, e.g., \cite{Dudal:2018cli}. In the following, we provide the expression for the tree-level propagator of the gauge field in the LVRGZ model in standard linear covariant gauges (meaning that we set $\zeta_{\mu\nu}=\mu^2 = 0$). We adopt the simplification in which $\kappa_{\mu\nu\alpha\beta} = 0$ and $a_{\mu} \to v_\mu$, in order to compare with the results of \cite{Santos:2014lfa}. The propagator is parameterized as follows,
\begin{eqnarray}
&&\langle A^a_{\mu} (p)A^b_{\nu} (-p)\rangle = \delta^{ab} \Bigg[a(p,v)\delta_{\mu\nu}+b(p,v){p_\mu p_{\nu}}\nonumber\\
&+&c(p,v)p_{\alpha} v_{\beta}\epsilon_{\mu\nu\alpha\beta}+d(p,v){v_{\mu}v_{\nu}} + {e(p,v)}\Sigma_{\mu\nu}\Bigg]\,,\nonumber\\
\label{Eq.SecTreeLevelGFP.1}
\end{eqnarray}
with $\Sigma_{\mu\nu} = v_{\mu}p_\nu + v_\nu p_\mu $. The corresponding form factors $a (p,v),\ldots , e(p,v)$ are given by
\begin{equation}
a(p,v) = \frac{\Lambda^2}{\Lambda^4-4\left[v^2 p^2 - (v\cdot p)^2\right]}\,,
\label{eq.a}
\end{equation}
\begin{equation}
b(p,v) = - \frac{\Xi+4\frac{v^2}{\Lambda^2}-\frac{(v\cdot p)^2}{p^2}\Omega\left(\Xi - \frac{v^2}{p^2}\right)}{\Lambda^2+ p^2\Xi - \frac{(v\cdot p)^2}{p^2}}a(p,v)\,,
\label{eq.b}
\end{equation}
\begin{equation}
c(p,v) = \frac{2i}{\Lambda^2}a(p,v)\,,
\end{equation}

\begin{equation}
d(p,v) = \Omega \,a(p,v)\,,
\end{equation}

\begin{equation}
e(p,v) = -\frac{v\cdot p}{p^2}d(p,v)\,,
\end{equation}
with
\begin{equation}
\Lambda^2 = p^2 + m^2 + \frac{2Ng^2 \gamma^4}{p^2+M^2}\,,
\end{equation}

\begin{equation}
\Xi = \frac{1}{\alpha} - 1 -\frac{m^2}{p^2} + \frac{(v\cdot p)^2}{p^4} - \frac{2Ng^2\gamma^4}{p^2 (p^2+M^2)}\,,
\end{equation}
and
\begin{equation}
\Omega =  -\frac{1+4\frac{p^2}{\Lambda^2}}{\Lambda^2+ v^2-\frac{(v\cdot p)^2}{p^2}}\,.
\end{equation}
As a consistency check, we take all the mass parameters as well as the Lorentz-violating structures to zero. This automatically eliminates the form factors $c(p,v)$, $d(p,v)$, and $e(p,v)$. Moreover, $\Lambda^2 = p^2$ and $\Xi = 1/\alpha - 1$. Hence,
\begin{equation}
\langle A^a_{\mu} (p)A^b_{\nu} (-p)\rangle_{\rm YM} = \delta^{ab}\left[\frac{1}{p^2}\mathcal{P}^{T}_{\mu\nu}(p) + \frac{\alpha}{p^2}\mathcal{P}^{L}_{\mu\nu}(p)\right]\,,
\label{consistcheck.1}
\end{equation}
with
\begin{equation}
\mathcal{P}^{T}_{\mu\nu}(p) = \delta_{\mu\nu} - \frac{p_\mu p_\nu}{p^2}\,,
\label{consistcheck.2}
\end{equation}
and
\begin{equation}
\mathcal{P}^{L}_{\mu\nu}(p) = \frac{p_\mu p_\nu}{p^2}\,,
\label{consistcheck.3}
\end{equation}
being the transverse and the longitudinal projectors, respectively. From eq.\eqref{consistcheck.1}, one sees that the standard tree-level propagator of the gauge field in YM theories quantized in standard linear covariant gauges is recovered. Another consistency check is to take all the LV-terms to zero and analyze the resulting propagator, i.e.,
\begin{equation}
\langle A^a_{\mu} (p)A^b_{\nu} (-p)\rangle_{\rm RGZ} = \delta^{ab}\left[\mathcal{D}(p^2)\mathcal{P}^{T}_{\mu\nu}(p) + \frac{\alpha}{p^2}\mathcal{P}^{L}_{\mu\nu}(p)\right]\,,
\label{consistcheck.4}
\end{equation}
with
\begin{equation}
\mathcal{D}(p^2) = \frac{p^2+M^2}{(p^2+m^2)(p^2+M^2)+2Ng^2\gamma^4}\,.
\label{consistcheck.5}
\end{equation}
The expression \eqref{consistcheck.4} together with \eqref{consistcheck.5} correspond to the tree-level propagator of the RGZ action in standard linear covariant gauges. We point out that both expressions \eqref{consistcheck.1} and \eqref{consistcheck.4} share the same longitudinal component. This is not a mere coincidence but a result that is exact thanks to the Ward identity associated with the equation of motion of the Nakanishi-Lautrup field. In fact, the longitudinal component in standard linear covariant gauges is exact and equals its tree-level expression, i.e., it does not renormalize.
This can be verified by noticing that in standard linear covariant gauges, the following Ward identity holds,
\begin{equation}
\frac{\delta \Gamma}{\delta b^a} = \partial_\mu A^a_\mu - \alpha b^a\,.
\label{consistcheck.6}
\end{equation}
It can be rewritten as 
\begin{equation}
J^{a}_{(b)}(x) = \partial_\mu \frac{\delta \EuScript{W}}{\delta J^a_\mu (x)} - \alpha \frac{\delta \EuScript{W}}{\delta J^a_{(b)}(x)}\,,
\label{consistcheck.7}
\end{equation}
with $J^{a}_{(b)}(x)$ representing the source coupled to the $b$-field and $J^a_\mu (x)$, the source coupled to $A^a_\mu$. Acting with $\delta/\delta J^b_\nu (y)$ on \eqref{consistcheck.7} leads to
\begin{equation}
0 = \partial^{x}_\mu \frac{\delta^2 \EuScript{W}}{\delta J^a_\mu (x) \delta J^b_\nu (y)} - \alpha \frac{\delta^2 \EuScript{W}}{\delta J^a_{(b)} (x) \delta J^b_\nu (y)}\,.
\label{consistcheck.8}
\end{equation}
Hence,
\begin{equation}
\partial^{x}_\mu\langle A^a_\mu (x) A^b_\nu  (y)\rangle = \alpha \langle b^a (x) A^b_\nu (y) \rangle \,.
\label{consistcheck.9}
\end{equation}
On the other hand, acting with $\delta/\delta J^b_{(b)}(y)$ on \eqref{consistcheck.7} yields
\begin{equation}
\delta^{ab}\delta(x-y)= \partial^x_\mu \frac{\delta^2\EuScript{W}}{J^a_\mu (x) J^b_{(b)}(y)} - \alpha \frac{\delta^2\EuScript{W}}{\delta J^a_{(b)}(x)\delta J^b_{(b)}(y)} \,.
\label{consistcheck.11}
\end{equation}
Hence,
\begin{equation}
\delta^{ab}\delta(x-y)= \partial^x_\mu \langle A^a_\mu (x) b^b (y) \rangle - \alpha \langle b^a(x) b^b (y)\rangle \,.
\label{consistcheck.12}
\end{equation}
In a BRST-invariant theory, the propagator $\langle b^a(x) b^b (y)\rangle$ vanishes exactly since it can be cast as a BRST-exact correlation function. Hence, in Fourier space,
\begin{equation}
\langle A^a_\mu (x) A^b_\nu (y) \rangle = \int \frac{{\rm d}^4 p}{(2\pi)^4}{\rm e}^{-ip\cdot (x-y)} \langle A^a_\mu A^b_\nu \rangle (p)\,,
\label{consistcheck.13}
\end{equation}
and
\begin{equation}
\langle A^a_\mu (x) b^b (y) \rangle = \int \frac{{\rm d}^4 p}{(2\pi)^4}{\rm e}^{-ip\cdot (x-y)} \langle A^a_\mu \,b^b \rangle (p)\,.
\label{consistcheck.14}
\end{equation}
Equation \eqref{consistcheck.12} allows for the exact derivation\footnote{One has to be careful here because, in principle, it is possible to have Lorentz-violating structures containing $v_\mu$. Nevertheless, the tree-level evaluation does not show such a structure and since this is an exact derivation, they should not appear at any order.} of the mixed propagator, i.e.,
\begin{equation}
\langle A^a_\mu b^b\rangle (p) = i\delta^{ab}\frac{p_\mu}{p^2}\,.
\label{consistcheck.15}
\end{equation}
Plugging eq.\eqref{consistcheck.15} in \eqref{consistcheck.9} in Fourier space, one obtains
\begin{equation}
p_\mu \langle A^a_\mu A^b_\nu \rangle (p) = \alpha\frac{p_\nu}{p^2}\,.
\label{consistcheck.16}
\end{equation}
This reveals a strong property of standard linear covariant gauges. Even considering the LVRGZ action (i.e., the LVYM action supplemented with the elimination of Gribov copies and the account for the generation of condensates), the logitudinal component of the gluon propagator is given by \eqref{consistcheck.16} and is exact. Hence, the gauge-field propagator in the LVRGZ theory is transverse in the Landau gauge $\alpha \to 0$ since \eqref{consistcheck.16} vanishes. This can be explicitly checked by taking the expressions eq.\eqref{eq.a} and \eqref{eq.b} and writing
\begin{equation}
a(p,v)\mathcal{P}^T_{\mu\nu}(p) + \left(a(p,v) + p^2 b(p,v)\right)\mathcal{P}^L_{\mu\nu}(p)\,.
\label{consistcheck.17}
\end{equation}
Keeping the LV parameters together with the RGZ ones, one can show that 
\begin{equation}
a(p,v) + p^2 b(p,v) = \frac{\alpha}{p^2}\,,
\label{consistcheck.18}
\end{equation}
which is an important consistency check of our computations.

There is another consistency check to be investigated which corresponds to keep the LV-terms and remove the mass-parameters that are present in the RGZ framework with the exception of the parameter $m^2$. Such a choice of parameters should provide the same propagator as the one obtained in \cite{Santos:2014lfa} in the Landau gauge limit, i.e., $\alpha \to 0$. This comparison requires that the following parameters introduced in \cite{Santos:2014lfa} 
\begin{eqnarray}
\Delta\Big|_{SS} &=& 6\alpha + 4\beta \\
\Omega\Big|_{SS} &=& 4\beta\,,
\label{consistcheck.19}
\end{eqnarray}
where we employ the subscript $SS$ to denote the quantities defined in \cite{Santos:2014lfa} are mapped as follows 
\begin{eqnarray}
\beta &=& - \frac{1}{4}\,,\\
\Delta\Big|_{SS} v^2 &=& \left(6\alpha -1\right)v^2 \equiv m^2\,,\\
\Omega\Big|_{SS} &=& 1\,.
\label{consistcheck.20}
\end{eqnarray}
An explicit inspection shows that our results for vanishing $(\gamma^2,M^2)$ are in agreement with the gauge-field propagator reported in \cite{Santos:2014lfa}, leading to another important consistency check. Thus, the expression reported in \eqref{Eq.SecTreeLevelGFP.1} corresponds to a consistent expression for the LVRGZ theory in standard linear covariant gauges. The expression defined by eq.\eqref{Eq.SecTreeLevelGFP.1} can be inspected and one can search for the underlying pole-structure. This is is not the purpose of the present work, but it is known that the presence of massive RGZ-parameters typically lead to complex poles violating reflection positivity, see, e.g., \cite{Dudal:2010tf,Cucchieri:2011ig,Dudal:2018cli}. As for the massive extension of LVYM theories, the pole-structure was investigated in \cite{Santos:2016uds}. A complete analysis involving all parameters, i.e., LV-parameters and RGZ-parameters is left for future investigation. 

%-------------------------------------------------------
\section{Conclusions \label{Sec:Conc}}
%-------------------------------------------------------

In this work, we have revisited the issue raised in \cite{Santos:2014lfa} regarding the dynamical mass generation of gauge fields in LVYM theories in the Landau gauge. In particular, we have proposed a massive extension of LVYM theories that is compatible with gauge invariance and coincides with the action proposed in \cite{Santos:2014lfa} in the Landau gauge. However, unlike the result obtained in \cite{Santos:2014lfa}, the model introduced in the present paper can be extended to gauges beyond the Landau choice and therefore is compatible with BRST symmetry. Such a property ensures that correlation functions of gauge-invariant operators do not depend on the choice of gauge parameters as the ones introduced in this paper for the sake of concreteness. This settles an important open question regarding the analysis performed \cite{Santos:2014lfa}: It is possible to introduce a mass-like term for the gauge fields in LVYM theories in harmony with gauge invariance/BRST symmetry. 

Secondly, we have argued that the mass generation observed in \cite{Santos:2014lfa} is a direct consequence of applying Symanzik's procedure together with the algebraic renormalization framework. By enlarging the LV-theory into a Lorentz-symmetric one, the authors couple sources to terms that are not quadratic on the fields, i.e., are not terms that break Lorentz invariance in a soft manner. This is an intrinsic property of the non-Abelian nature of the theory, since the Chern-Simons-like term as well as the aether term are quadratic on the fields in the Abelian limit and, hence, could be made Lorentz invariant by the promotion of the fixed background vector and tensor to sources. Such a difference entails the generation of terms that are quadratic on the gauge field coupled to the Lorentz-symmetry restoring sources which under their physical limit generate mass-like terms for the gauge field. In this sense, one starts with LVYM which is gauge invariant. Afterwards, Landau gauge is fixed and Lorentz symmetry is recovered by the promotion of background vectors and tensors to sources. Finally, the algebraic renormalization prescription is followed and the final counterterm contains structures that couple the symmetry-restoring sources and gauge fields. By choosing the physical value of the symmetry-restoring sources, mass terms are generated and the resulting theory breaks BRST invariance. This is slightly weird since it might suggest that the theory has an anomalous Slavnov-Taylor identity. In principle, in the Landau gauge, such a breaking (partially) vanishes on-shell and perhaps is harmless, but a similar analysis in linear covariant gauges (and corresponding extensions) would generate a proliferation of gauge-parameter dependent correlation functions of gauge-invariant quantities. Hopefully, the model here introduced solves this issue albeit it was not proved explicitly that it is renormalizable in a general linear covariant gauge, a task left to future work due to its complexity. Hence, a consistent massive model extension (with respect to gauge invariance) of LVYM can be introduced and can be used as an effective model for computations in the context of infrared LVYM theories. Yet it is our expectation that explicit computations within LVYM or the analysis under the algebraic renormalization framework of LVYM will not generate mass terms explicitly. 

Having addressed the nature of the mass term in LVYM theories, we have worked out a fully BRST-invariant elimination of infinitesimal Gribov copies. This has generated a LVRGZ theory and the gauge-field propagator was computed in standard linear covariant gauges. Consistency tests were performed confirming the viability of the model. 

This work opens up several different avenues to be explored in the context of LVYM theories. As a first step, one should reconsider the algebraic renormalization of LVYM theories (in linear covariant gauges, for instance) without using Symanzik's method in order to explicitly verify the absence or not of mass generation. As a second step, a proof of renormalizability of the massive extension is desirable in order to make sure if the mass-parameters $m^2$ and $a_\mu$ must be tied to the already present LV parameters. Furthermore, one should perform a comprehensive analysis of the renormalizability properties of the LVRGZ action and investigate if new condensates with LV-structures can be generated or not. All those questions are interesting on their own and are really helpful to understand the deep role played by Lorentz invariance in non-Abelian gauge theories and in its infrared regime. A comprehensive analysis of the different phases that such a theory can develop thanks to the presence of the LV-parameters is achievable due to the pole structure of the gauge-field propagator. Finally, an explicit computation in LVYM theories up to second-order in the LV-parameters in desirable in order to verify any sort of mass generation from a different perspective. Those are open problems that deserve to be explored to enrich our understanding of structural aspects of non-Abelian gauge theories and Lorentz symmetry. 

%-------------------------------------------------------
\section*{Acknowledgments}
%-------------------------------------------------------

ADP is thankful to A. Lehum, A. Petrov and R. F. Sobreiro for clarifications and discussions. ADP acknowledges CNPq under the grant PQ-2 (312211/2022-8), FAPERJ under the “Jovem Cientista do Nosso Estado” program (E26/202.800/2019 and E-26/205.924/2022).

%%%%%%%%%%%%%%%%%%%%%%
%%%%%%APPENDICES%%%%%%%%%%
%%%%%%%%%%%%%%%%%%%%%%
%\newpage

%\appendix

%\begin{widetext}

%\end{widetext}

%\newpage
\bibliography{refs}

\end{document}